\documentstyle[aps,preprint]{revtex}

\begin{document}
\draft
\title{Spinor Decomposition of $SU(2)$ Gauge Potential and The Spinor Structures of
Chern-Simons and Chern Density}
\author{Yi-shi Duan$^1$, Xin Liu$^1$\thanks{%
Author to whom correspondence should be addressed. Electronic address:
liuxin@lzu.edu.cn} and Li-bin Fu$^2$}
\address{$^1${\it Institute of Theoretical Physics, Lanzhou University, }\\
{\it Lanzhou 730000, P. R. China}\\
$^2${\it LCP, Institute of Applied Physics and Computational Mathematics, }\\
{\it P.O. Box 8009(26), Beijing 100088, P.R. China}}
\maketitle

\begin{abstract}
In this paper, the decomposition of $SU(2)$ gauge potential in terms of
Pauli spinors is studied. Using this decomposition, the spinor strutures of
the Chern-Simons form and the Chern density are obtained. Furthermore, by
these spinor structures, the knot quantum number of non-Abelian gauge theory
is discussed, and the second Chern number is characterized by the Hopf
indices and the Brouwer degrees of $\phi $-mapping.
\end{abstract}


\section{Introduction}

In recent years the decomposition theory of gauge potential is playing a
more and more important role in theoretical physics and mathematics. Since
the decomposition theory reveals the inner structure of gauge potential, it
inputs the topological and other information to the gauge potential (i.e.
the connection of principal bundle), and establishes a direct relationship
between differential geometry and topology of gauge field. From this
viewpoint much progress has been made by other authors \cite{Faddeev,cho0}
and by us, such as the decomposition of $U(1)$ gauge potential and the $U(1)$
topological quantum mechanics, the decomposition of $SO(N)$ spin connection
and the structure of GBC topological current, and the decomposition of $%
SU(N) $ connection and the effective theory of $SU(N)$ QCD, etc. \cite
{DuanGe,DZh,DLnpb,DFjmp,DLSU(N)}.

The Pauli spinor is the fundamental representation of $SU(2)$ gauge field.
In Sect.II of this paper, the spinor decomposition of $SU(2)$ gauge
potential is studied. In Sect.III, using this decomposition, we obtain the
spinor structure expression of Chern-Simons form, by which the knot quantum
number of non-Abelian gauge field is studied. In comparison with the
representation of the $SU(2)$ gauge potential, the expression of knot
quantum number in terms of spinor is more direct and precise in non-Abelian $%
SU(2)$ gauge theory. In Sect.IV, the spinor structure of $SU(2)$ Chern
density is obtained. By making use of the $\phi $-mapping topological
current theory, the Chern density is expressed as $\delta (\vec{\phi})$.
Therefore, the zero points of $\phi $ field are characterized by the Hopf
indices ($\beta _j$) and Brouwer degrees ($\eta _j$) of $\phi $-mapping, and
the second Chern number, which is directly related to the Euler
characteristic through the top Chern class on $4$-dimensional manifold, is
characterized by $\beta _j$ and $\eta _j$.

\section{The Spinor Decomposition of $SU(2)$ Gauge Potential}

It is well known that in $SU(2)$ gauge field theory for spinor
representation $\Psi $, the covariant derivative of $\Psi $ is defined as 
\begin{equation}
D_\mu \Psi =\partial _\mu \Psi -\frac 1{2i}A_\mu ^a\sigma _a\Psi ,
\label{covder}
\end{equation}
where 
\begin{equation}
{\bf A}={\bf A}_\mu dx^\mu =\frac 1{2i}A_\mu ^a\sigma _adx^\mu  \label{su2au}
\end{equation}
is the $SU(2)$ gauge potential (connection), and $T_a=\frac 1{2i}\sigma _a$ $%
(a=1,2,3)$ are the $SU(2)$ generators with $\sigma _a$ the Pauli matrices.
The complex conjugate of $D_\mu \Psi $ is 
\begin{equation}
D_\mu ^{\dagger }\Psi ^{\dagger }=\partial _\mu \Psi ^{\dagger }+\frac 1{2i}%
\Psi ^{\dagger }A_\mu ^a\sigma _a.  \label{c.c}
\end{equation}
The $SU(2)$ gauge field tensor is given by 
\begin{equation}
{\bf F}_{\mu \nu }=\partial _\mu {\bf A}_\nu -\partial _\nu {\bf A}_\mu -[%
{\bf A}_\mu ,{\bf A}_\nu ],  \label{boldFuv}
\end{equation}
where 
\begin{equation}
{\bf F}=\frac 12{\bf F}_{\mu \nu }dx^\mu \wedge dx^\nu ,\;{\bf F}_{\mu \nu }=%
\frac 1{2i}F_{\mu \nu }^a\sigma _a.  \label{Fuvcliff}
\end{equation}

To complete the decomposition of $SU(2)$ gauge potential, multiplying Eq.(%
\ref{covder}) with $\Psi ^{\dagger }\sigma _b$ and Eq.(\ref{c.c}) with $%
\sigma _b\Psi $ respectively and using 
\begin{equation}
\sigma _a\sigma _b+\sigma _b\sigma _a=2\delta _{ab}I,  \label{orthog}
\end{equation}
one can easily find 
\begin{equation}
A_\mu ^a=\frac i{\Psi ^{\dagger }\Psi }(\Psi ^{\dagger }\sigma _a\partial
_\mu \Psi -\partial _\mu \Psi ^{\dagger }\sigma _a\Psi )-\frac i{\Psi
^{\dagger }\Psi }(\Psi ^{\dagger }\sigma _aD_\mu \Psi -D_\mu ^{\dagger }\Psi
^{\dagger }\sigma _a\Psi ).  \label{Amjua}
\end{equation}

Since any $2\times 2$ Hermitian matrix $X$ can be expressed in terms of
Clifford basis $(I,\;\vec{\sigma}):$%
\begin{equation}
X=\frac 12Tr(X)I+\frac 12Tr(X\sigma _a)\sigma _a,  \label{Mtrace}
\end{equation}
from Eqs.(\ref{su2au}), (\ref{Amjua}) and (\ref{Mtrace}) we can obtain 
\begin{equation}
{\bf A}_\mu ={\bf a}_\mu +{\bf b}_\mu ,  \label{bigaudecom}
\end{equation}
where 
\begin{eqnarray}
{\bf a}_\mu =\frac 1{\Psi ^{\dagger }\Psi }(\partial _\mu \Psi \Psi
^{\dagger }-\Psi \partial _\mu \Psi ^{\dagger }) &&-\frac 1{2\Psi ^{\dagger
}\Psi }Tr(\partial _\mu \Psi \Psi ^{\dagger }-\Psi \partial _\mu \Psi
^{\dagger })I,  \label{amu} \\
{\bf b}_\mu =-[\frac 1{\Psi ^{\dagger }\Psi }(D_\mu \Psi \Psi ^{\dagger
}-\Psi D_\mu ^{\dagger }\Psi ^{\dagger }) &&-\frac 1{2\Psi ^{\dagger }\Psi }%
Tr(D_\mu \Psi \Psi ^{\dagger }-\Psi D_\mu ^{\dagger }\Psi ^{\dagger })I].
\label{bmu}
\end{eqnarray}
It is easy to prove that ${\bf a}_\mu $ and ${\bf b}_\mu $ satisfies the
gauge transformation and the vectorial transformation respectively: 
\begin{eqnarray}
{\bf a}_\mu ^{^{\prime }} &=&S{\bf a}_\mu S^{\dagger }+\partial _\mu
SS^{\dagger },  \label{autran} \\
{\bf b}_\mu ^{^{\prime }} &=&S{\bf b}_\mu S^{\dagger },  \label{butran}
\end{eqnarray}
where $S^{\dagger }=S^{-1}\;(S\in SU(2)),$ hence ${\bf A}_\mu $ satisfies
the required $SU(2)$ gauge transformation \cite{Nash,Schwarz} 
\begin{equation}
{\bf A}_\mu ^{^{\prime }}=S{\bf A}_\mu S^{\dagger }+\partial _\mu
SS^{\dagger }.  \label{Autrans}
\end{equation}
{\em Therefore Eq.(\ref{bigaudecom}) with Eqs.(\ref{amu}) and (\ref{bmu}) is
just the spinor decomposition of }$SU(2)${\em \ gauge potential.}

\section{The Spinor Structures of The Chern-Simons Form and The Knot Quantum
Number}

Let $M$ be a compact oriented $4$-dimensional manifold, on which there is an
open cover $\{U,\;V,\;W,\;...\}$ with transition function $S_{uv}$
satisfying 
\begin{equation}
S_{uu}=1,\;S_{uv}^{-1}=S_{vu},\;S_{uv}S_{vw}S_{wu}=1.\;(U\cap V\cap W\neq
\emptyset )
\end{equation}
On the principal bundle $P(\pi ,\;M,\;SU(2)),$ the Chern-Simons $3$-form is
defined as \cite{ChnSim,Jackiw,Nash} 
\begin{equation}
\Omega =\frac 1{8\pi ^2}Tr({\bf A}\wedge d{\bf A}-\frac 23{\bf A}\wedge {\bf %
A}\wedge {\bf A}),  \label{chnsimform}
\end{equation}
i.e. 
\begin{equation}
\Omega =-\frac 1{16\pi ^2}\epsilon ^{\mu \nu \lambda }[A_\mu ^a\partial _\nu
A_\lambda ^a-\frac 13\epsilon ^{abc}A_\mu ^aA_\nu ^bA_\lambda ^c]d^3x.
\label{2terms}
\end{equation}
This leads to the second Chern class \cite{Chncls} 
\begin{eqnarray}
c_2(P) &=&d\Omega ,  \label{chnweil} \\
c_2(P) &=&\frac 1{8\pi ^2}Tr({\bf F}\wedge {\bf F})=\rho (x)d^4x,
\label{c2(P)}
\end{eqnarray}
where $\rho (x)$ is the $SU(2)$ Chern density.

In the traditional decomposition theory of gauge potential (including
Riemann geometry) \cite{DuanGe,DLnpb}, always using the parallel field
condition 
\begin{equation}
D_\mu \Psi =0,  \label{parellel}
\end{equation}
the solution is then ${\bf b}_\mu =0,$ and $A_\mu ^a$ can be solved in terms
of $\Psi $ 
\begin{equation}
A_\mu ^a=a_\mu ^a=\frac i{\Psi ^{\dagger }\Psi }(\Psi ^{\dagger }\sigma
_a\partial _\mu \Psi -\partial _\mu \Psi ^{\dagger }\sigma _a\Psi ).
\label{psinot=1}
\end{equation}

In the above text the spinor $\Psi $ is a $2\times 1$matrix 
\begin{equation}
\Psi =\left( 
\begin{array}{l}
\phi ^0+i\phi ^1 \\ 
\phi ^2+i\phi ^3
\end{array}
\right) ,\;  \label{psiphi}
\end{equation}
where $\phi ^a\;(a=0,1,2,3)$ are real functions, $\phi ^a\phi ^a=\left\|
\phi \right\| ^2=\Psi ^{\dagger }\Psi $. For simplicity, we introduce a unit
vector $n^a\;(a=0,1,2,3)$%
\begin{equation}
n^a=\frac{\phi ^a}{\left\| \phi \right\| },\;n^an^a=1.  \label{ndef}
\end{equation}
Obviously the zero points of $\phi ^a$ are just the singular points of $n^a$%
. And a normalized spinor $\Psi _n$ is introduced: 
\begin{equation}
\Psi _n=\frac 1{\sqrt{\Psi ^{\dagger }\Psi }}\Psi =\left( 
\begin{array}{l}
n^0+in^1 \\ 
n^2+in^3
\end{array}
\right) .  \label{psinewdef}
\end{equation}
In following, without making mistakes, we can still use the signal ''$\Psi $%
'' instead of ''$\Psi _n$'' to denote the normalized spinor. Thus Eq.(\ref
{psinot=1}) becomes 
\begin{equation}
A_\mu ^a=i(\Psi ^{\dagger }\sigma _a\partial _\mu \Psi -\partial _\mu \Psi
^{\dagger }\sigma _a\Psi ).  \label{psinorm}
\end{equation}

Then we can use Eqs.(\ref{2terms}) and (\ref{psinorm}) to study the spinor
structure of $\Omega $. Noticing that the Pauli matrix elements satisfy the
formulas 
\begin{eqnarray}
\sigma _a^{\alpha \beta }\sigma _a^{\alpha ^{\prime }\beta ^{\prime }}
&=&2\delta _{\alpha \beta ^{\prime }}\delta _{\alpha ^{\prime }\beta
}-\delta _{\alpha \beta }\delta _{\alpha ^{\prime }\beta ^{\prime }},
\label{Pauliele1} \\
\epsilon _{abc}\sigma _a^{\alpha \beta }\sigma _b^{\alpha ^{\prime }\beta
^{\prime }}\sigma _c^{\alpha ^{\prime \prime }\beta ^{\prime \prime }}
&=&-2i(\delta _{\alpha \beta ^{\prime }}\delta _{\alpha ^{\prime }\beta
^{\prime \prime }}\delta _{\alpha ^{\prime \prime }\beta }-\delta _{\alpha
\beta ^{\prime \prime }}\delta _{\alpha ^{\prime \prime }\beta ^{\prime
}}\delta _{\alpha ^{\prime }\beta }),  \label{Pauliele2}
\end{eqnarray}
we arrive at 
\begin{equation}
\Omega =-\frac 1{4\pi ^2}\Psi ^{\dagger }d\Psi \wedge d\Psi ^{\dagger
}\wedge d\Psi .  \label{chnsimnew}
\end{equation}
This is just {\em the spinor structure of Chern-Simons }$3${\em -form of }$%
SU(2)$ {\em gauge field theory.}

The above spinor structure of Chern-Simons form can be applied in studying
the knot quantum number of non-Abelian gauge theory. The quantum number of
Faddeev-Niemi knot is given by the integration in $3$-dimension \cite{knot} 
\begin{equation}
Q_{FN}=\frac 1{32\pi ^2}\int \epsilon _{ijk}C_iH_{jk}d^3x,\;(i,j,k=1,2,3)
\label{FNknot}
\end{equation}
where $H_{ij}$ is an Abelian gauge field tensor 
\begin{equation}
H_{ij}=-\vec{m}\cdot (\partial _i\vec{m}\times \partial _j\vec{m})=\partial
_iC_j-\partial _jC_i,\;(\vec{m}\cdot \vec{m}=1)
\end{equation}
and $\vec{m}$ is the non-linear $\sigma $-model field; in $SU(2)$ gauge
field $m^a=\Psi ^{\dagger }\sigma ^a\Psi \;(a=1,2,3).$ Here $Q_{FN}\in \pi
_3(S^2)\;(\pi _3(S^2)={\bf Z})$, so it is the Hopf invariant.

It is known that \cite{Niemi2} 
\begin{equation}
\frac 14\epsilon _{ijk}C_iH_{jk}=\epsilon _{ijk}Tr({\bf A}_i\partial _j{\bf A%
}_k-\frac 23{\bf A}_i{\bf A}_j{\bf A}_k),  \label{ab-nonab}
\end{equation}
therefore the Faddeev-Niemi model is related to the non-Abelian $SU(2)$
gauge field theory 
\begin{equation}
Q=\int \Omega =\frac 1{8\pi ^2}\int \epsilon _{ijk}Tr({\bf A}_i\partial _j%
{\bf A}_k-\frac 23{\bf A}_i{\bf A}_j{\bf A}_k).  \label{nabelknt}
\end{equation}
and $Q\in \pi _3(S^3)\;(\pi _3(S^3)={\bf Z).}$ Since $\pi _3(S^3)=\pi
_3(S^2) $ \cite{Nash}, Eqs.(\ref{FNknot}) and (\ref{nabelknt}) give the same
knot quantum number, which is just the vacuum number of the $SU(2)$ gauge
potential \cite{cho,Cho2,BaalWipf}.

In this paper, using the spinor expression of Chern-Simons form (\ref
{chnsimnew}), from Eq. (\ref{nabelknt}) the knot quantum number can be
directly expressed as 
\begin{equation}
Q=\int \Omega =-\frac 1{4\pi ^2}\int \epsilon _{ijk}\Psi ^{\dagger }\partial
_i\Psi \partial _j\Psi ^{\dagger }\partial _k\Psi d^3x.  \label{choknot}
\end{equation}
In comparison with Eq.(\ref{nabelknt}), the expression of knot quantum
number $Q$ in terms of spinor is obviously more direct and precise in
non-Abelian $SU(2)$ gauge field theory.

\section{The Spinor Structure of Chern Density and The Inner Structure of
The Second Chern Number}

From Eqs.(\ref{chnweil}) and (\ref{chnsimnew}) we can obtain {\em the spinor
structure of the second Chern class}: 
\begin{equation}
c_2(P)=-\frac 1{4\pi ^2}d\Psi ^{\dagger }\wedge d\Psi \wedge d\Psi ^{\dagger
}\wedge d\Psi ,  \label{2ndchnclsnew}
\end{equation}
and {\em the spinor structure of }$SU(2)${\em \ Chern density}: 
\begin{equation}
\rho (x)=-\frac 1{4\pi ^2}\epsilon ^{\mu \nu \lambda \rho }\partial _\mu
\Psi ^{\dagger }\partial _\nu \Psi \partial _\lambda \Psi ^{\dagger
}\partial _\rho \Psi .  \label{chndensnew}
\end{equation}

In terms of the unit vector $n^a,$ the Chern density $\rho (x)$ (Eq.(\ref
{chndensnew})) can be expressed as \cite{DFjmp} 
\begin{equation}
\rho (x)=\frac 1{12\pi ^2}\epsilon ^{\mu \nu \lambda \rho }\epsilon
_{abcd}\partial _\mu n^a\partial _\nu n^b\partial _\lambda n^c\partial _\rho
n^d.  \label{chndensn}
\end{equation}
By making use of our $\phi $-mapping topological current theory \cite
{DuanGe,DZh,DLnpb,DFjmp,DLSU(N),p-brane,DuanSLAC}, we can use Eq.(\ref{ndef}%
) and the Green function relation in $\phi $-space 
\begin{equation}
\frac{\partial ^2}{\partial \phi ^a\partial \phi ^a}(\frac 1{\left\| \phi
\right\| ^2})=-4\pi ^2\delta ^4(\vec{\phi})
\end{equation}
to reexpress Chern density $\rho (x)$ in a $\delta $-function form \cite
{p-brane,DuanSLAC} 
\begin{equation}
\rho (x)=\delta ^4(\vec{\phi})D(\frac \phi x),  \label{chndensdelta}
\end{equation}
where $D(\phi /x)$ is the Jacobi determinant 
\begin{equation}
\epsilon ^{abcd}D(\frac \phi x)=\epsilon ^{\mu \nu \lambda \rho }\partial
_\mu \phi ^a\partial _\nu \phi ^b\partial _\lambda \phi ^c\partial _\rho
\phi ^d.
\end{equation}

The implicit function theory shows that \cite{Goursat}, under the regular
condition $D(\phi /x)\neq 0$, the general solutions of 
\begin{equation}
\phi ^a(x^0,x^1,x^2,x^3)=0\;(a=0,1,2,3)
\end{equation}
can be expressed as $N$ isolated points 
\begin{equation}
x^\mu =x_j^\mu .\;(\mu =0,1,2,3;\;j=1,...,N)  \label{Npoints}
\end{equation}
In $\delta $-function theory \cite{Schouton}, one can prove 
\begin{equation}
\delta ^4(\vec{\phi})=\sum_{j=1}^N\frac{\beta _j\delta ^4(x^\mu -x_j^\mu )}{%
\left| D(\phi /x)\right| _{x_j^\mu }},
\end{equation}
where the positive integer $\beta _j$ is the Hopf index of $\phi $-mapping.
In topology it means that when the point $x^\mu $ covers the neighborhood of
the zero point $x_j^\mu $ once, the vector field $\phi ^a$ covers the
corresponding region in $\phi $-space $\beta _j$ times. Introducing the
Brouwer degree of $\phi $-mapping 
\begin{equation}
\eta _j=\frac{D(\phi /x)}{\left| D(\phi /x)\right| _{x_j^\mu }}=sign[D(\phi
/x)]_{x_j^\mu }=\pm 1,  \label{eta}
\end{equation}
Eq.(\ref{chndensdelta}) can be expressed as 
\begin{equation}
\rho (x)=\sum_{j=1}^N\beta _j\eta _j\delta ^4(x^\mu -x_j^\mu ).
\label{rouBjHj}
\end{equation}
Eq.(\ref{rouBjHj}) directly shows that the Chern density does not vanish
only at the $N$ $4$-dimensional zero points of $\phi ^a$, i.e. the singular
points of $n^a$, which are characterized by the Hopf indices $\beta _j$ and
the Brouwer degrees $\eta _j$ of $\phi $-mapping.

Furthermore, when integrating the second Chern class, one obtains the second
Chern number: 
\begin{equation}
C_2=\int c_2(P)=\int \rho (x)d^4x=\sum_{j=1}^N\beta _j\eta _j.  \label{C2num}
\end{equation}
Since the base manifold $M$ is $4$-dimensional, the second Chern class $%
c_2(P)$ is just the top Chern class on $P$; on the other hand, there is a
direct relation between the top Chern class and the Euler class \cite{Eguchi}
\begin{equation}
c_2(P)=e(E),  \label{topchncls}
\end{equation}
where $E$ is a real vector bundle which is the real counterpart of complex
vector bundle $P$, and $e(E)$ is the Euler class on $E$. Therefore the Euler
characteristic, which is just the sum of indices of zero points of vector
field $\phi ^a$ on $M$, is obtained through the Gauss-Bonnet theorem 
\begin{equation}
\chi (M)=\int e(E)=\sum_{j=1}^N\beta _j\eta _j.  \label{Eulernum}
\end{equation}
So the indices of zero points of $\phi ^a$ field can be composed of the
topological numbers $\beta _j$ and $\eta _j$.

At last there are two points which should be stressed. Firstly, besides in
this paper, the spinor decomposition of $SU(2)$ gauge potential can also be
applied in studying the $U(1)$ field tensor in $SU(2)$ gauge field.
Secondly, when the self-dual condition \cite{Nash} 
\begin{equation}
{\bf F}_{\mu \nu }^{*}={\bf F}_{\mu \nu }\;({\bf F}_{\mu \nu }^{*}=\frac 12%
\epsilon _{\mu \nu \lambda \rho }{\bf F}_{\lambda \rho })  \label{instant}
\end{equation}
is satisfied, the corresponding zero points of $\phi ^a$ field on ${\bf R}^4$
are just the instantons, so their topological numbers can also be studied.
These two points will be detailed in our other papers.

\section{Acknowledgment}

This work was supported by the National Natural Science Foundation and the
Doctor Education Fund of Educational Department of the People's Republic of
China.

\end{document}